\documentclass[%
reprint,
superscriptaddress,
 amsmath,amssymb,
 aps,
]{revtex4-1}

\usepackage{graphicx}
\usepackage{dcolumn}
\usepackage{bm}

\usepackage{setspace} 
\usepackage{subcaption} 
\usepackage{float} 
\usepackage{xspace} 
\usepackage{xfrac} 
\usepackage[hidelinks]{hyperref} 
\hypersetup{
colorlinks,
citecolor=black,
filecolor=black,
linkcolor=black,
urlcolor=black
} 
\usepackage{xcolor} 

\newcommand{\wn}{cm$^{-1}$\xspace}
\newcommand{\Cs}{CsPbBr$_3$\xspace}
\newcommand{\MA}{CH$_3$NH$_3$PbBr$_3$\xspace}
\newcommand{\ma}{CH$_3$NH$_3^+$\xspace}

\begin{document}


\title{Interplay between organic cations and inorganic framework and incommensurability in hybrid lead-halide perovskite \MA}

\author{Yinsheng Guo}
\affiliation{Department of Chemistry, Columbia University, New York, NY 10027, USA}
\author{Omer Yaffe}
\affiliation{Department of Materials and Interfaces, Weizmann Institute of Science, Rehovot, 76100, Israel}
\author{Daniel W. Paley}
\affiliation{Department of Chemistry, Columbia University, New York, NY 10027, USA}
\affiliation{Columbia Nano Initiative, Columbia University, New York NY 10027}
\author{Alexander N. Beecher}
\affiliation{Department of Chemistry, Columbia University, New York, NY 10027, USA}
\author{Trevor D. Hull}
\affiliation{Department of Chemistry, Columbia University, New York, NY 10027, USA}
\author{Guilherme Szpak}
\affiliation{Departamento de Fisica, Universidade Federal de Minas Gerais, 30123-970 Belo Horizonte, Brazil}
\affiliation{Department of Chemistry, Columbia University, New York, NY 10027, USA}
\author{Jonathan S. Owen}
\affiliation{Department of Chemistry, Columbia University, New York, NY 10027, USA}
\author{Louis E. Brus}
\affiliation{Department of Chemistry, Columbia University, New York, NY 10027, USA}
\author{Marcos A. Pimenta}
\email{mpimenta@fisica.ufmg.br}
\affiliation{Departamento de Fisica, Universidade Federal de Minas Gerais, 30123-970 Belo Horizonte, Brazil}
\affiliation{Department of Chemistry, Columbia University, New York, NY 10027, USA}

\date{\today}

\begin{abstract}
Organic-inorganic coupling in the hybrid lead-halide perovskite is a central issue in rationalizing the outstanding photovoltaic performance of these emerging materials.
Here we compare and contrast the evolution of structure and dynamics of the hybrid \MA and the inorganic \Cs lead-halide perovskites with temperature, using Raman spectroscopy and single-crystal X-ray diffraction.
Results reveal a stark contrast between their order-disorder transitions, abrupt for the hybrid whereas smooth for the inorganic perovskite.
X-ray diffraction observes an intermediate incommensurate phase between the ordered and the disordered phases in \MA.
Low-frequency Raman scattering captures the appearance of a sharp soft mode in the incommensurate phase, ascribed to the theoretically predicted amplitudon mode.
Our work highlights the interaction between the structural dynamics of organic cation \ma and the lead-halide framework, and unravels the competition between tendencies of the organic and inorganic moieties to minimize energy in the incommensurate phase of the hybrid perovskite structure.
\end{abstract}

\maketitle


Organic-inorganic hybrid lead-halide perovskites have emerged as a promising class of new generation photovoltaic materials\cite{Gratzel2014, Green2014a,Berry2015}, showing excellent electronic properties and outstanding power conversion efficiencies\cite{Zhou2014a}. 
Fundamental insight of the structural dynamics holds the key to understanding their unique electronic properties, 
and is the subject of intensive theoretical and experimental investigations.\cite{Yaffe2017, Sendner2016, Leguy2016, Brivio2015a, Mattoni2016,Mosconi2016} 
Methylammonium cation (\ma, MA), the most frequently incorporated A site cation in recent photovoltaic applications, possesses a large dipole and exhibits dynamic orientational disorder at room temperature.\cite{Poglitsch1987,Mashiyama2007} 
The orientational degrees of freedom of the anisotropic organic cation have been suggested to be responsible for the excellent electronic properties of the hybrid perovskites. \cite{Liu2015,Ma2015,Quarti2015}.
Yet the inorganic lead-halide framework is the actual optically and electronically active component, on which charge carriers reside.
A gap of knowledge exists on how the structural dynamics of MA and lead-halide framework couple with each other despite intensive investigations.

In this Letter, we focus on unravelling the interplay between organic MA cations and the inorganic lead-halide framework, and highlight the unique signatures of such coupling.
Here we report drastically different phase transformation behaviors between hybrid and inorganic lead-halide perovskites.
Using low-frequency Raman scattering, we observe that there is an intermediate structural phase in \MA between the orthorhombic and tetragonal phases. 
With single-crystal X-ray diffraction, we conclude that the intermediate state to be an incommensurate phase, and we follow the evolution of the incommensurate modulation wave vector as a function of temperature.
A newly activated Raman mode of the incommensurate phase is assigned to the vibration in the amplitude of the structural modulation (amplitudon mode).
The amplitudon shows soft mode behavior, revealing the displacive component in the phase transition of hybrid lead-halide perovskites often considered to be of order-disorder type.
The incommensurability in the hybrid perovskite is a direct consequence of the organic-inorganic coupling, and is absent in the all-inorganic counterpart. 

\begin{figure*}
\centering
\includegraphics[width=6.5in]{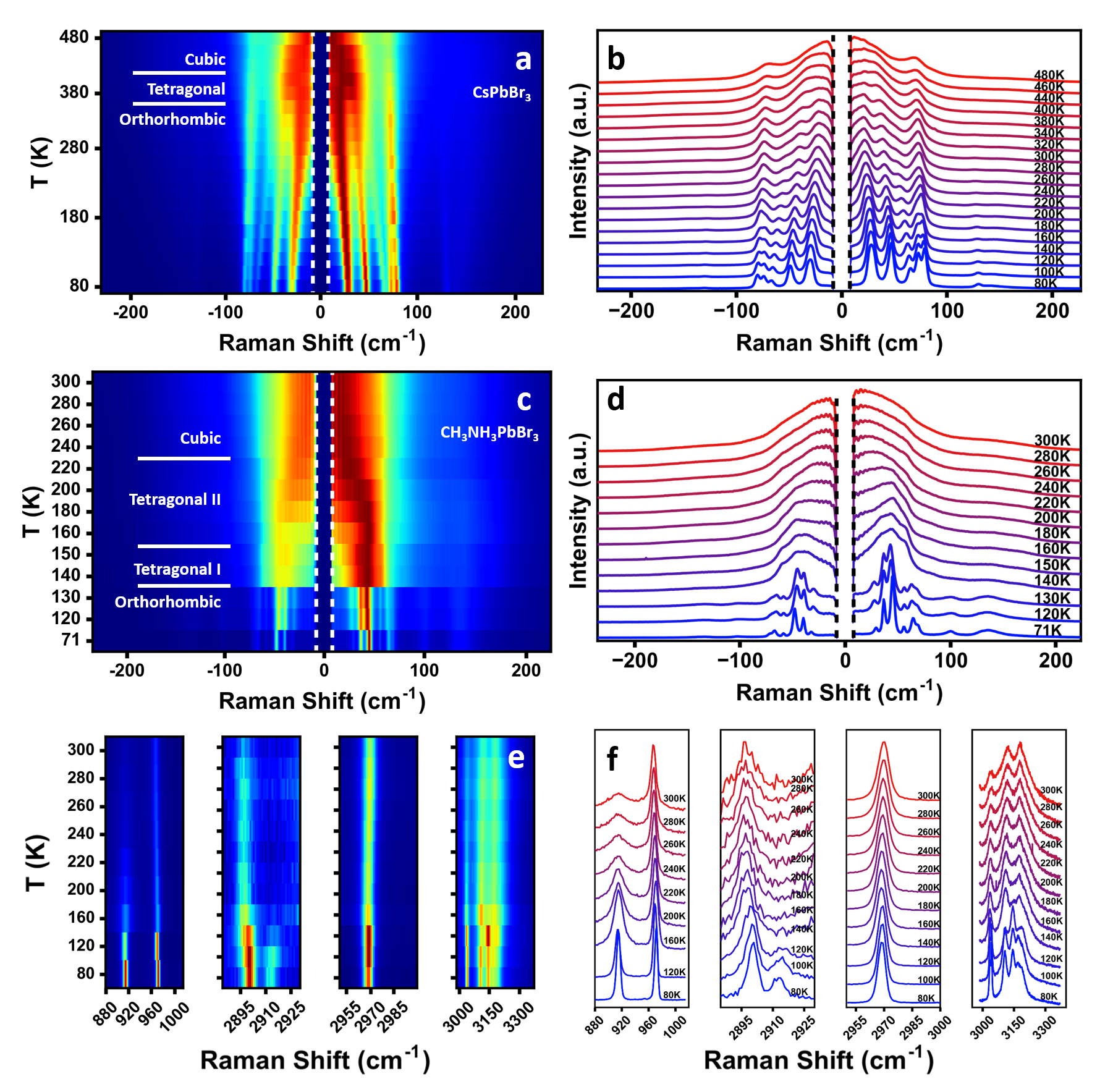} 
\caption{\bf Temperature dependent Raman spectra of hybrid and inorganic lead-halide perovskite crystals.
(a) color plot and (b) line plot of low-frequency Raman spectra of \Cs.
(c) color plot and (d) line plot of low-frequency Raman spectra of \MA.
(e) color plot and (f) line plot of mid- and high-frequency Raman spectra of \MA, 
including \ma internal modes such as 
NH$_3$ wagging at 910 \wn, 
C-N stretching at 969 \wn, 
C-H stretching at 2890 - 2970 \wn,
and N-H stretching around 3100 \wn.
Horizontal white lines in (a) and (c) indicate commensurate phase transition temperatures reported in previous studies.\cite{Poglitsch1987,Mashiyama2007,Stoumpos2013a,Hirotsu1974}
Raman spectra are normalized at each temperature, except for (e) to show the intensity drop of \ma internal modes across the orthorhombic-tetragonal phase transition.
}
\label{fig_spec_Tdep} 
\end{figure*}

\paragraph*{Different temperature dependence of the structural dynamics between hybrid and inorganic lead-halide perovskites}
Figure \ref{fig_spec_Tdep} shows the Raman spectra of \Cs and \MA perovskite crystals, in a broad range of temperatures across the phase transitions.
Vibrational modes naturally separate into different frequency regions according to the nature of their nuclear motion.
Collective framework modes and local Pb-Br vibrations reside at low-frequency (THz)\cite{Sendner2016, La-o-vorakiat2016,Yettapu2016} regions (Figure 1 a-d), while MA internal modes reside at mid- and high-frequency regions (Figure 1 e, f).
Comparing the frameworks modes' evolution with temperature, it is immediately clear that the phase transition behaviors are qualitatively different between the inorganic and hybrid lead-halide perovskites.
\Cs shows a gradual transition through the sequence of structural phases (Figure \ref{fig_spec_Tdep}a and b).
The Raman modes continuously broaden with temperature and the transitions appear to be of second order nature.
Unlike \Cs, at the orthorhombic-tetragonal transition, \MA shows a remarkably abrupt change from discrete modes to a continuous central peak and broad vibrational bands (Figure \ref{fig_spec_Tdep}c and d). 
The transition temperatures agree with previously reported XRD values, which are indicated by white lines in Figure \ref{fig_spec_Tdep}a.\cite{Poglitsch1987,Mashiyama2007,Stoumpos2013a,Hirotsu1974} 

In light of these apparent differences, and to understand the unique interaction of MA and the lead-halide network, we measured the Raman spectra at higher frequencies, corresponding to internal \MA Raman modes (Figure \ref{fig_spec_Tdep}e and f). 
The peak at {$\approx$} 320 \wn, observed in both bromide crystals (supporting information), appears to correspond to a 480 \wn peak in CH$_3$NH$_3$PbCl$_3$.\cite{Maalej1997} 
The frequency ratio here is about 1.5, in agreement with what would be expected for a simple harmonic oscillator model using only halide masses. 
The fact that this 320 \wn peak exists both in \MA and \Cs, and that the frequency ratio agrees with harmonic oscillators with halide masses, suggest that this mode is related to halide motion, and not the CH$_3$ torsion as previously suggested.\cite{Maalej1997}
Peaks at 910 \wn, 969 \wn, in the 2890-2970 \wn range, and around 3100 \wn appear only in \MA (Figure \ref{fig_spec_Tdep}c), and are assigned to NH$_3$ wagging, C-N stretch, C-H stretch and N-H stretch, respectively.\cite{Gray1957}
These internal \MA modes show an abrupt broadening and intensity decrease at the same temperature ( {$\approx$} 150 K) where the abrupt orthorhombic-tetragonal phase transition occurs in the low-frequency Raman spectra, indicating that the phase transition is coupled to the disordered motion of the \MA.\cite{Mattoni2015}

\begin{figure*}
\centering
\includegraphics[height=7in]{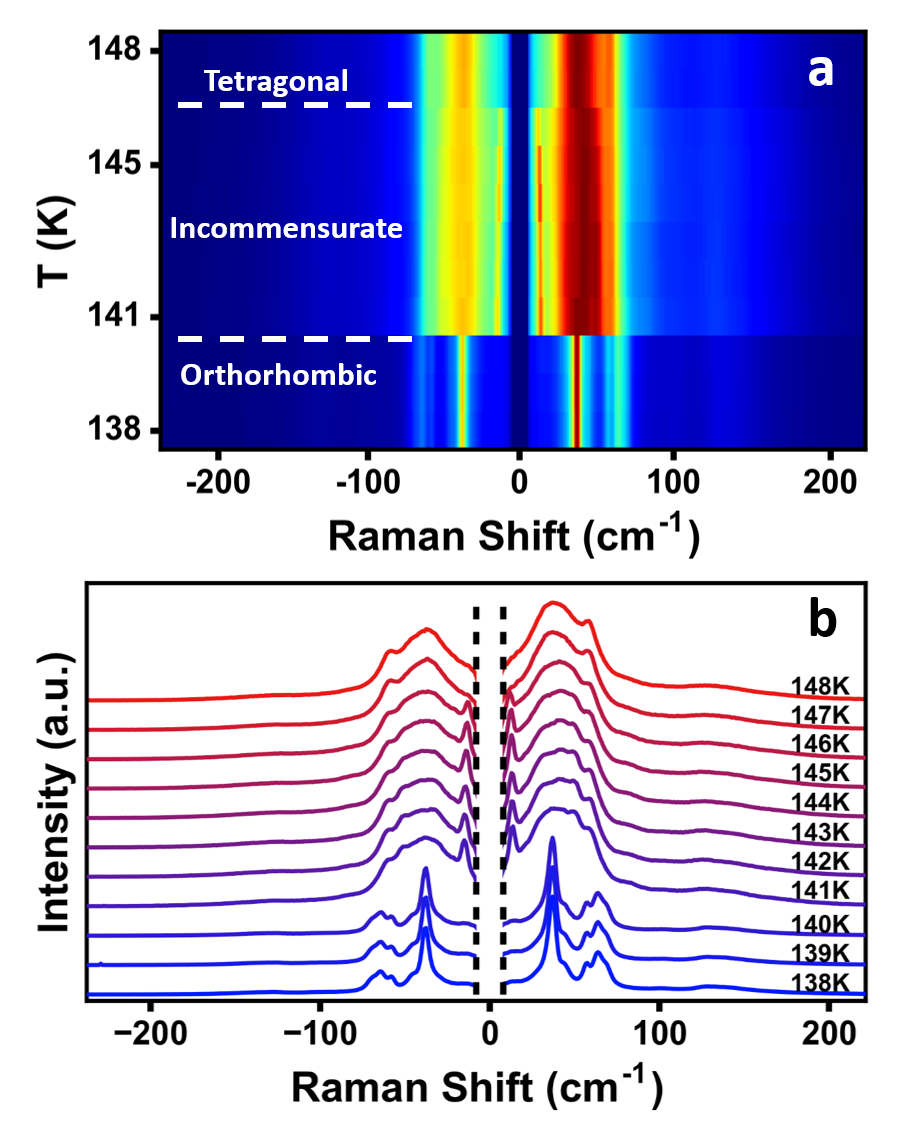} 
\caption{\bf Temperature dependent low frequency Raman spectra of the orthorhombic, incommensurate, and tetragonal phases of \MA. (a) color plot of spectra above, in, and below the incommensurate phase. (b) line plot of spectra at each temperature.}
\label{fig_spec} 
\end{figure*}

We now focus on the temperature range across the abrupt phase transition in \MA between the low temperature ordered phase and high temperature disordered phase. 
As shown in Figure \ref{fig_spec}, a sharp low frequency mode around 13 cm$^{-1}$ appears in the low frequency spectra below 146 K and disappears at the orthorhombic phase, below 140 K. 
This result shows the existence of an intermediate phase within a temperature window of about 5 K, between the orthorhombic phase and tetragonal phase.

\begin{figure*}
\centering
\includegraphics[width=\textwidth]{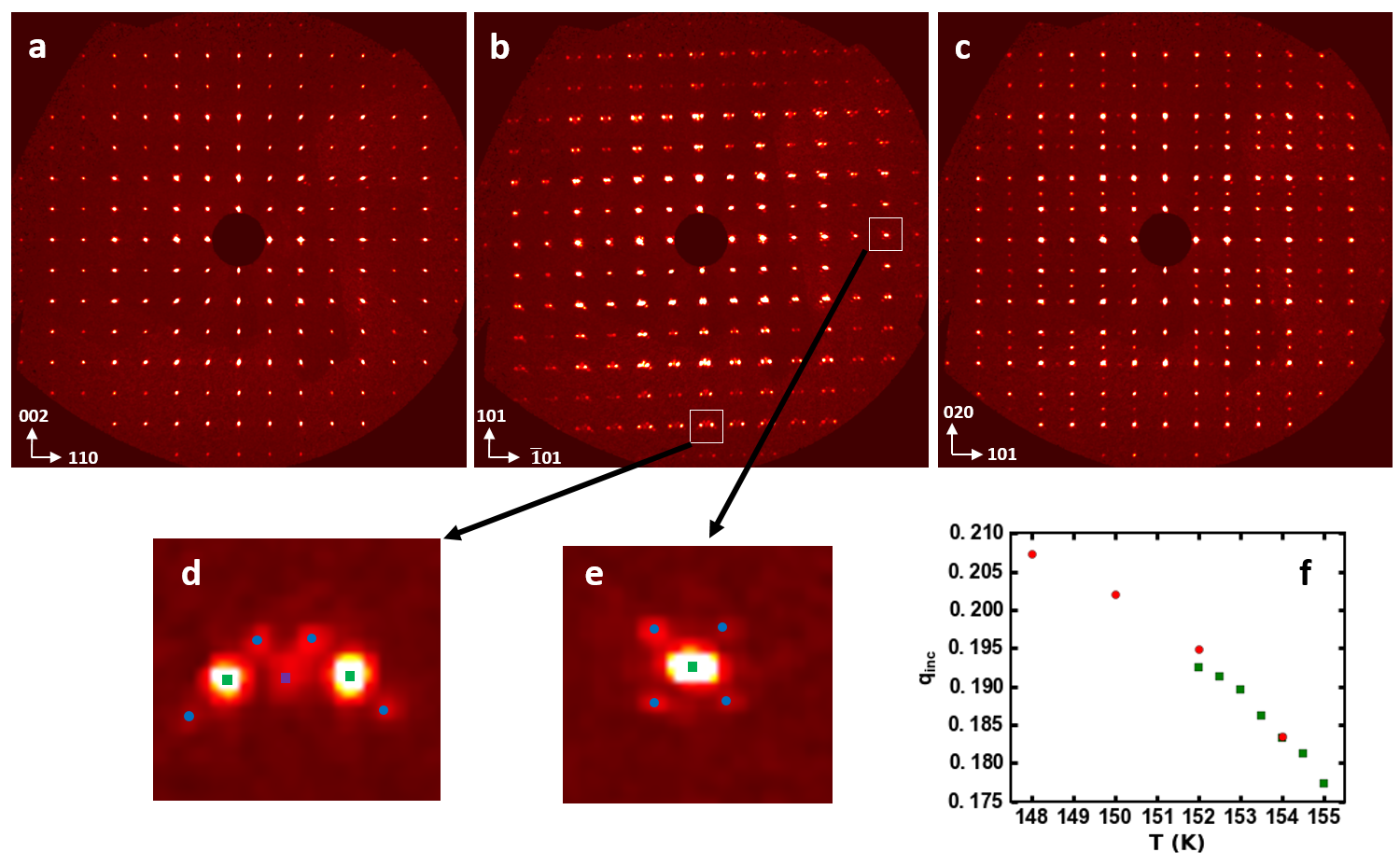} 
\caption{\bf 
Single crystal X-ray diffraction of \MA crystals. 
Reciprocal lattice reconstructions for \MA at (a) 160 K; (b) 150 K; (c) 140 K. 
The views are along a principal direction of the parent (cubic) structure and are labeled for the conventional settings of I4/mcm, Imma and Pnma respectively. 
(d) and (e) Splitting of reflections by twinning and modulation.
Reflection points are taken from marked boxes in (b), and are magnified by the same scale.
Green squares are main reflections (m=0) for two twin-related domains.
Blue dots are satellite reflections (m=00$\pm$1).
The unindexed spot marked by a purple square is a reflection from a further minor twin domain.
The incommensurate modulation wave vectors $q_{inc}$ are defined by the blue dot - green square pairs, whose angle and magnitude are identical in (d) and (e).
(f) Evolution of the incommensurate modulation wave vector $q_{inc}$ with temperature. 
Red dots and green squares represent data from two different measurements.}
\label{fig_xrd} 
\end{figure*}

\paragraph*{The incommensurate structure of the intermediate phase}
Single crystal X-ray diffraction measurements were performed to determine the intermediate phase lattice structure.
Reciprocal lattice reconstructions for \MA at 160 K, 150 K and 140 K are shown in figure \ref{fig_xrd}. 
The views are along a principal direction of the parent (cubic) structure. 
The X-ray diffraction spots at 160 K shown in figure \ref{fig_xrd}a are labeled for the conventional settings of the tetragonal I4/mcm structure (one antiphase octahedral tilt\cite{Glazer1972}). 
On the other hand, results obtained at 140 K shown in Fig \ref{fig_xrd}c indicate that the ordered low temperature phase exhibits an orthorhombic Pnma symmetry, in agreement with a previous result by neutron powder diffraction by Swainson et al.\cite{swainson2015} 

The X-ray diffraction map at 150 K shown in figure \ref{fig_xrd}b exhibits additional satellite reflections at a fixed wave vectors {\bf q} away from each main reflection. 
Interestingly, these new satellites cannot be described by rational numbers of the reciprocal unit cell vectors. 
As plotted in figure \ref{fig_xrd}f, the wave vector evolves from 0.1773 c* at 155 K to 0.2073 c* at 148 K, as plotted in figure \ref{fig_xrd}f. 
This is an unambiguous evidence of an incommensurate modulated structure along the c-axis of the hybrid perovskite crystal.
An incommensurate structure occurs when a crystalline lattice superimposes with an additional distortional wave and the periodicities of the two structure cannot be expressed with a rational-number ratio. 
The incommensurate modulation generates the additional satellite reflection at a fixed wave vector q$_{inc}$. 
Our observation is consistent with the recent neutron scattering study by Swainson and coworkers, reporting an intermediate phase in \MA speculated to be an incommensurate structure.\cite{swainson2015}

The average structure of the incommensurate phase is orthorhombic, and belongs to the body-centered space group Imma, which is also a standard perovskite structure described by two equal antiphase tilts.\cite{Glazer1972} 
Notably, neither of these two phase transitions (tetragonal to incommensurate and incommensurate to orthorhombic) occur through an apparent group-subgroup relation, and therefore they are first-order transitions as previously described by Knop et al.\cite{Knop1990}

The incommensurate phase is also twinned by a two-fold rotation around hkl=(-1 0 1). 
The different splittings of reflections due to modulation and twinning are easily distinguishable because the incommensurate q-vector is independent of the scattering vector h while the twin-related splitting is hkl-dependent. 
The difference is illustrated in Figures \ref{fig_xrd}d and \ref{fig_xrd}e. 

\begin{figure*}
\includegraphics[width=\textwidth]{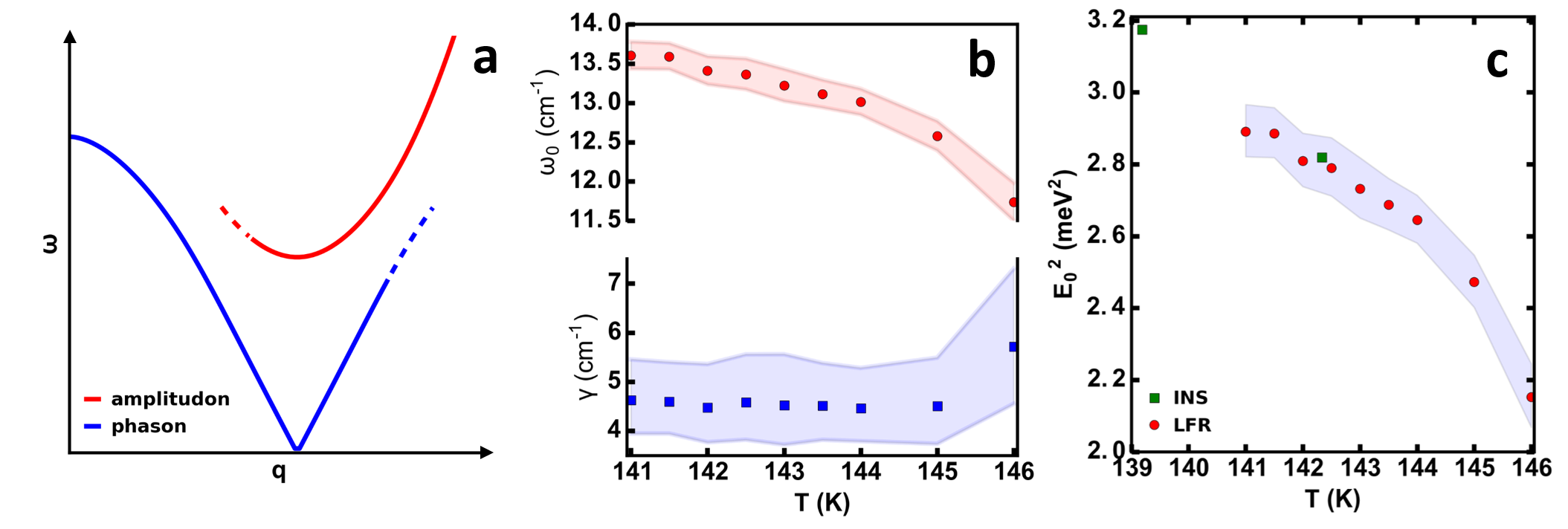}
\caption{\bf Characteristics of the soft amplitudon mode within the incommensurate phase of \MA. 
(a) Schematic phonon dispersion curves of the amplitudon and phason branches. Adapted from reference \citenum{Cummins1990}.
(b) Temperature dependent line shift (upper panel) and linewidth (lower panel) of the amplitudon mode in the incommensurate phase.
(c) Soft mode shifting behavior. 
LFR: red circles plot the low frequency Raman measurement results. 
INS: green squares plot lowest energy data points from the inelastic neutron scattering measurement\cite{swainson2015}.}
\label{fig_softmode}
\end{figure*}

\paragraph*{Lattice dynamics of incommensurate structures}
Before analyzing the Raman results in the intermediate incommensurate phase, let us first discuss the theoretical predictions for the lattice dynamics of incommensurate structures. 
One common mechanism that drives the formation of incommensurate structure is a decending (optic) phonon branch interacting with a low-lying (acoustic) phonon branch.
These two modes have different symmetries at the Brillouin zone center and boundaries, but share an identical symmetry at all intermediate points. 
As the upper branch decends and approaches the lower branch, the anti-crossing interaction mixes and repels the two phonon modes. 
After the diagonalization of the two-mode-coupling matrix, symmetric and antisymmetric combinations are produced. 
Both modes can be considered normal modes of the incommensurate structure for small $q$.\cite{Dvorak1978}
The in-phase mode describes the amplitude modulation of the static distortion, thus is termed amplitudon, whereas the out-of-phase mode describes the phase modulation of the frozen incommensurate wave, thus is termed phason.  
A schematic of typical dispersion relations of the amplitudon and phason branches are shown in figure \ref{fig_softmode}a. 
Amplitudon modes have been previously observed in the incommensurate phases of other compounds \cite{Cummins1990,Klein1983}, but phason modes are often elusive to Raman scattering observation. 
Generally the phason mode frequency approaches zero towards zone center, while the damping is largely independent on momentum.\cite{Cummins1990}.
Therefore the phason mode is usually at a low energy very close to the strong Rayleigh scattering, and often too overdamped for clear identification. 

\paragraph*{Raman spectra of the incommensurate phase and amplitudon soft mode}
The representative low frequency Raman spectra of the tetragonal, incommensurate and orthorhombic phases are shown in Figure \ref{fig_spec}. 
Upon entering the incommensurate phase, the discrete normal modes in the orthorhombic phase become significantly overdamped and broadened into a continuum feature. 
Additionally, a strong peak abruptly appears at $\approx$ 13 \wn. 
This new mode has a very narrow linewidth, in stark contrast with the broad band centered around 50 \wn.
As the \MA crystal is heated up into the tetragonal phase, the sharp peak disappears and the broad band remains. 

Since the wavelength of the modulation cannot be related to the crystalline lattice constant with a rational ratio, the translational symmetry is broken along the modulation direction.
This symmetry breaking relaxes the crystal momentum conservation and thus Raman selection rules. 
Many Raman processes away from the $\Gamma$ point in the Brillouin zone, forbidden in the original phase, are now allowed. 
The activation of scattering processes across the Brillouin zone results in the spectral continuum in the incommensurate phase, which corresponds to the broad band centered around 50 \wn in Figure \ref{fig_softmode}. 

The sharp mode at $\approx$ 13 \wn is only observed in the incommensurate phase, indicating that it is a normal mode of the incommensurate modulated structure. 
The appearance of a spectrally well-defined new normal mode in the incommensurate phase agrees with the theoretical prediction of the amplitudon mode, which is associated with the vibration of the incommensurate modulation amplitude.
The temperature dependence of the frequency of this sharp mode plotted in Figure \ref{fig_softmode}b.
This mode softens as temperature increases with a thermal coefficient of  $\approx$0.26 \wn/K, an order of magnitude larger than the thermal coefficient of $\approx$0.02 \wn/K of the orthorhombic phase normal modes. 
This rapid softening behavior is drastically different compared to the peaks observed in the orthorhomic phase. 
Interestingly, both the frequency and its strong temperature dependence of this mode match very well with the quickly softening mode observed within the low temperature orthorhombic phase using inelastic neutron scattering\cite{swainson2015}, shown in Figure \ref{fig_softmode}c.
In the orthorhombic phase, this softening mode was associated purely with the tilting of the PbBr$_3$ octahedral framework, likely away from Brillouin zone center.
Across the commensurate-incommensurate transition, this soft lead-halide octahedral tilting mode couples and mixes with the MA ordering.
The organic-inorganic interaction drives phonon condensation at a k point within the Brillouin zone of the orthorhombic phase; 
this condensation point then becomes the center of the incommensurate phase Brillouin zone.
By tracking the evolution of this octahedral tilting mode into the incommensurate phase amplitudon, 
we ellucidate the identity of the lead-halide framework vibration that participates in the organic-inorganic interaction.

The observation of soft mode behavior also highlights the displacive component in the phase transformation between the orthorhombic and tetragonal phases in lead-halide perovskites. 
Generally, the coexistence of both order-disorder and displacive components have now been shown for many materials previously considered to be single component prototypes.\cite{Zalar2003,Mountstevens2005}
For hybrid perovksites, much discussion in the literature focused on the order-disorder component of the MA rotation, and the displacive component received relatively little attention.\cite{swainson2015} 
Therefore, the observation of the amplitudon soft mode in the low frequency Raman scattering complement the understanding of the structural dynamics in hybrid lead-halide perovskites.
It is worth to comment that low frequency Raman scattering is uniquely suited to measure this amplitudon mode in the incommensurate structure.
Observation was achieved beyond the most softened phonon state of the orthorhombic phase seen by inelastic neutron scattering, 
thanks to the superior spectral resolution and sharp spectral cutoff of the excitation afforded by the optical measurements. 

\paragraph*{Frustrated interplay between the dynamics of the organic and inorganic parts of the hybrid perovskite}
The study of structurally incommensurate phases has widely attracted research interest and is an ever-growing field.\cite{Dvorak1978, Cummins1990, Jorio1998}
Mechanistic elucidation of incommensurate structures offer insight on a wide variety of research frontiers, ranging from fundamental understanding of exotic electronic and spin structures and properties,\cite{Li2014c,Smirnov2009} to practical engineering of optical applications.\cite{Zhao2017}
Incommensurability is always associated with the concept of frustration, namely a structural compromise that results from two incompatible interactions competing with each other.\cite{swainson2005} 
In our low frequency Raman scattering observations, incommensurate phases exist in the hybrid perovskite \MA, but is absent in the inorganic perovskite \Cs. 
This comparison indicates that the competing interaction between of the dipolar organic cations and the perovskite framework is responsible for the incommensurate structure.
Such competition occurs when the molecular dipolar coupling and the lead-halide octahedra tilts are both thermally activated and their ordering processes interact with each other.

Structural instabilities have already been predicted\cite{Lehner1982, Huang2014, DaSilva2015}, and recently observed\cite{Beecher2016} for halide perovskites.
CsSnX$_3$ (X=I, Br, Cl) has been recently investigated theoretically, as an inorganic homologue to the hybrid perovskites.\cite{Huang2014}
In the orthorhombic phase, all phonon modes calculated are well behaved with real and positive frequencies.
However, in the tetragonal and cubic phases imaginary frequencies are produced for some phonon branches, which indicate structural instability, and identifies these branches as soft phonon modes.
The condensation of soft phonon branches at the Brillouin zone edge drives the phase transition between commensurate structures.
Phonon modes associated with the tilting motion of the lead-halide octahedra can be found from specific phonon branches at the Brillouin zone edge k points.
However, in the hybrid perovskites, the softening of a lead-halide framework phonon branch could interact and compete with another ordering process, namely the dipolar coupling of the MA cations.
This interaction drives the condensation to occur at a low symmetry point within the Brillouin zone and away from the zone edge, leading to the incommensurate modulation.

The MA ion's behavior is abruptly different below and above the incommensurate phase, as shown in Figure \ref{fig_spec_Tdep}c.
At low temperatures below the incommensurate phase, the MA ions are orderly confined in the crystalline lattice, and MA ion Raman modes are strong and sharp. 
At high temperatures above the incommensurate phase, the MA ions are loosely bound, and rotate in the cubo-octahedral cage. 
The internal Raman modes of MA ion become weak and broadened.
The narrow incommensurate temperature window and the drastic difference of the MA behavior suggest the dipolar ordering of the organic cation is critically dependent on the available thermal energy. 
Other hybrid perovskite also show incommensurate phases, [(CH$_3$)$_4$N]GeCl$_3$ is a well studied compound\cite{Futterer1995} where the incommensurate structure arises from a distortion of a frozen transverse acoustic phonon. 

Our results allow us to conclude that the existence of this incommensurate phase is a manifest of the frustrated competition between the tendencies of the organic dipolar cations and inorganic frameworks to minimize their energies.
Specifically, at the incipient melting of the MA sublattice, weak long range MA ordering interacts with the softening octahedral tilting motion and results in the incommensurate phase.
As temperature rises, thermal energy overwhelms the MA dipolar ordering, which leads to the transition into the commensurate tetragonal structure.

In summary, we study the interplay between the organic cations and inorganic framework in \MA and report the unique signatures of the coupled structural dynamics.
The structural phase transition is abrupt for the hybrid lead-halide perovskites and is associated with the melting of dipolar organic cation, whereas it is continuous and occurs in a broad temperature range for the all-inorganic counterpart.
We observe for the hybrid perovskite \MA an intermediate phase between the ordered (orthorhombic) and disordered (tetragonal) phases. 
Single crystal X-ray diffraction confirms the that the intermediate phase is incommensurate and revealed its structural evolution with temperature.
We discover a new amplitudon mode of the incommensurate structure using low-frequency Raman scattering.
This amplitudon mode is related to the vibration of the modulation amplitude, and results from the interaction of structural dynamics between organic MA cations and inorganic lead-halide octahedral tilting.
The soft mode behavior reveals the displacive component of structural phase transition, which is mainly considered as order-disorder type in the hybrid lead-halide perovskites.

The incommensurability and the discovery of the new amplitudon mode reflect the competition to minimize energy between the incipient melting of the organic MA sublattice with the octahedral tilting motion in the inorganic lead-halide framework, and highlight the interplay between the dynamics of organic dipolar cations and the inorganic framework in hybrid lead-halide perovskite \MA.

\paragraph*{Acknowledgements}
Y.G. acknowledges financial support from the Keck foundation. 
O. Y. acknowledges funding by the FP7 People program under the project Marie Curie Grant No. IOF-622653.
M.A.P. acknowledges the financial support from the Brazilian agencies CNPq, CAPES, FAPEMIG and Brazilian Nanocarbon Institute of Science and Technology (INCT-Nanocarbono).
X-ray diffraction was performed in the Shared Materials Characterization Laboratory at Columbia University. 
Use of the Shared Materials Characterization Laboratory was made possible by funding from Columbia University.

\bibliography{LHP_inc}

\end{document}